# Research Protocol for the Google Health Digital Well-being Study

Daniel McDuff, Google Research
Andrew Barakat, Google Research
Ari Winbush, University of Oregon
Allen Jiang, Google Research
Felicia Cordeiro, Google Research
Ryann Crowley, University of Oregon
Lauren E. Kahn, University of Oregon
John Hernandez, Google Research
Nicholas B. Allen, University of Oregon

## Abstract

Background: The impact of digital device use on health and well-being is a pressing question to which individuals, families, schools, policy makers, legislators, and digital designers are all demanding answers. However, the scientific literature on this topic to date is marred by small and/or unrepresentative samples, poor measurement of core constructs (e.g., device use, smartphone addiction), and a limited ability to address the psychological and behavioral mechanisms that may underlie the relationships between device use and well-being.  A number of recent authoritative reviews have made urgent calls for future research projects to address these limitations. The critical role of research is to identify which patterns of use are associated with benefits versus risks, and who is more vulnerable to harmful versus beneficial outcomes, so that we can pursue evidence-based product design, education, and regulation aimed at maximizing benefits and minimizing risks of smartphones and other digital devices. We describe a protocol for a Digital Well-Being (DWB) study to help answer these questions.

Objective: The objectives of this study are to provide normative data on objective patterns of smartphone use. We aim to: (1) identify how patterns of smartphone use impact well-being, identify groups of individuals who show similar patterns of covariation between smartphone use and well-being measures across time, (2) examine sociodemographic and personality/mental health predictors and which patterns of smartphone use and well-being are associated with pre-post changes in mental health and functioning, (3) discover which non-device behavior patterns (e.g., sleep, physical activity, geographic movement, social interactions) mediate the association between device use and well-being, (4) identify & explore recruitment strategies to increase and improve the representation of traditionally under-represented populations, (5) provide a real-world baseline of observed stress, mood, insomnia, physical activity and sleep across a representative population.

Methods: This is a prospective, nonrandomized study to investigate patterns and relationships between digital device use, sensor based measures (including both behavioral and physiological signals), and self-reported measures of mental health and well-being. The study duration is four-weeks long per participant and includes passive sensing based on smartphone sensors, and optionally a wearable (Fitbit), for the complete four-week period.

Results: At the time of submission, the study infrastructure and app has been designed and built, the institutional review board of the University of Oregon has approved the study protocol and data collection is underway. Data from 4,182 enrolled and consented participants has been collected as of March 27, 2023.

Conclusions: The impact of digital devices on mental health and well-being raises important questions. The DWB Study is designed to help answer questions about how the association between patterns of smartphone use and well-being.

Trial Registration: Not applicable.

Keywords: Digital well-being, mobile, wearable, mental health, smartphones

# Introduction

The impact of digital device use on health and well-being is a pressing question to which individuals, families, schools, policy makers, legislators, and digital designers are all demanding answers (Haidt and Allen 2020). However, the scientific literature on this topic to date is marred by small and/or unrepresentative samples, poor measurement of core constructs (e.g., device use, smartphone addiction), and a limited ability to address the psychological and behavioral mechanisms that may underlie the relationships between device use and well-being. A number of recent authoritative reviews have made urgent calls for future research projects to address these limitations (e.g., (Girela-Serrano et al. 2022); (Fang et al. 2021; Odgers and Jensen 2020; Orben 2020)). The critical role of research is to identify which patterns of use are associated with benefits versus risks, and who is more vulnerable to harmful versus beneficial outcomes, so that we can pursue evidence-based product design, education, and regulation aimed at maximizing benefits and minimizing risks of smartphones and other digital devices.

Leveraging objective data from smartphones and wearables is needed to advance scientific research in this area. Physiological and behavioral biomarkers that can be measured by smartphones and wearables present a valuable way of identifying those at risk and measuring risk factors. Promising prior research has found that longitudinal smartphone and wearable data can be used to predict and, perhaps more importantly, complement existing clinical measures (Zhang et al. 2021; Wang et al. 2018; Fang et al. 2021). Indeed, the variables that can be captured by these devices can help us understand some of the most important determinants of health. For example, behavioral risk factors such as poor mental health, sleep, and physical activity are the greatest predictor of preventable illness and death (GBD 2016 Risk Factors Collaborators 2017), and nearly 8% of the US GDP is lost to mental disorders each year (Arias, Saxena, and Verguet 2022). However, mental and behavioral health conditions are complex and multi-faceted phenomena and measurement is non-trivial. Illness can manifest in a wide range of symptoms (Galatzer-Levy and Bryant 2013) that impact both physiological states and behavior. While there are examples of impressive longitudinal studies (Fang et al. 2021; Xu et al. 2023), the scientific literature on these topics is still dominated by studies with small and often unrepresentative samples, poor measurement of core constructs (e.g., device use), self-reported measures of behavior that may not correlate with objectively measured endpoints, and a limited ability to address the psychological and behavioral mechanisms that may underlie the relationships between device use and well-being. For example, some studies focus solely on college students (Xu et al. 2023) or medical interns (Fang et al. 2021). A number of recent authoritative reviews have made urgent calls for future research projects to address these limitations (Fang et al. 2021; Odgers and Jensen 2020; Orben 2020); (Prinstein, Nesi, and Telzer 2020).

To address this gap in the literature we have designed a large-scale, intensive longitudinal study (the Digital Well-being Study [DWB]) to analyze relationships between smartphone use, wearable and smartphone physiological and behavioral data, and self-reported measures of mental health and well-being. We will recruit up to 14,000 participants who opt-in to complete intake and outtake surveys, daily ecological assessments and enable sensing of features from their smartphone. In addition, at least 50% of the participants will wear a smartwatch (Fitbit). Each individual's participation in the study will be for 28 days.

The aims of the Digital Well-being Study are manifold. We aim to:
- Provide normative data on objective patterns of smartphone use, and describe how these patterns of use vary with age, gender, geography, and functioning/mental health.
- Identify how patterns of smartphone use impact well-being, and who is most vulnerable to harmful effects related to smartphone use.
- Identify groups of individuals who show similar patterns of covariation between smartphone use and well-being measures across time.
- Examine sociodemographic and personality/mental health predictors.
- Examine which patterns of smartphone use and well-being are associated with pre-post changes in mental health and functioning, and compare the predictive capacity of these objective measures to relevant self-report measures (e.g. smartphone addiction measures).
- Discover which non-device behavior patterns (e.g., sleep, physical activity, geographic movement, social interactions) mediate the association between device use and well-being.
- Identify & explore recruitment strategies to increase and improve the representation of traditionally under-represented populations.
- Provide a real-world baseline of observed stress, mood, insomnia, physical activity and sleep across a representative population.

Next, will describe the methods used in our data collection protocol and preliminary statistics on the demographic profile of the first cohort of subjects.

## Methods

### Study Design

The Digital Well-being Study is a prospective, nonrandomized study to investigate patterns and relationships between digital device use patterns, including sensor data from phones and wearables reflecting both behavioral and physiological processes, and self-reported measures of mental health and well-being. The study is four weeks long with passive sensing from a smartphone, and optionally a wearable (Fitbit), for the complete four-week period.

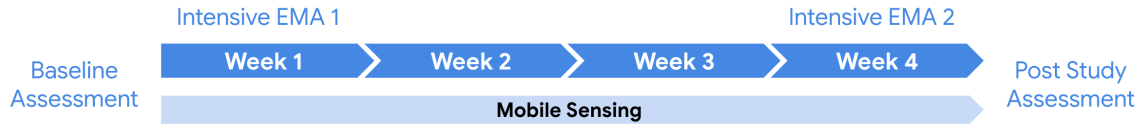

Fig 1. The design of the four week intensive digital sensing study. Mobile sensing comprises measurements from a smartphone and, optionally, a wrist-worn wearable.

### Recruitment

A key goal of the study is to explore methods to collect a representative study sample that reflects the relative proportions of various demographic subgroups, and their intersectionalities in adults of the US population. For example, according to the US Census Bureau, 51% are female and 44% adults are between 20 and 44 years of age. Among those who reported a single race, about 74% are Caucasian, 13% are African American, and 6% are Asian. A February 2021 Gallup poll reported that 5.6% of US adults identified themselves as lesbian, gay, bisexual, or transgender (LBGT). 86.7% said that they were heterosexual or straight, and 7.6% refused to answer. In the past, some subgroups have been traditionally underrepresented in health studies, e.g., females, African Americans, and those with a LGBTQIA+ orientation (Oh et al. 2015), and recent reviews have emphasized the importance of setting explicit recruitment goals in overcoming these biases (Oh et al. 2015; Cullen et al. 2023). As such a key aim of the study is to learn more about how to increase the participation of traditionally underrepresented groups in digital health studies.

To explore methods to overcome recruitment barriers of traditionally underrepresented subgroups and make sure the study sample reflects sufficient diversity, we will pursue a two-prong approach: (a) a primary-augmented two-wave metered recruitment schema and (b) multi-stakeholder engagement. 'Metered Recruitment' schema refers to the potential delay of eligible study participants in consent and enrollment (baseline characterization, and initiation into the monitoring portion of the study). The pace of enrollment will be adapted based on study and clinical bandwidths, and the distributions of demographic characteristics, with metering instituted by invitation to consent. The research team plans to begin the metered recruitment strategy once the study has achieved or has nearly achieved enrollment of n = 10,000. Depending on distribution of key demographic variables, once consent and appropriate permissions are obtained, enrollment and initiation of monitoring will commence. To help reduce the effect of seasonality and other external factors on the data collected in this study we are careful to design the strategy to recruit participants in a balanced manner across time and outside of periods containing major seasonal holidays.

Wave 1 (Open Enrollment with enriched recruitment): During the initial several weeks that the study is open, we anticipate enrollment of up to approximately 1,000 new participants per week (but will not limit to this pace). Enrollment will increase as infrastructure allows. An increase in consecutive enrollment of up to 10,000 new participants, who are eligible for study participation will be invited to begin the enrollment process. The study will monitor the progress in recruiting traditionally underrepresented groups. Recruitment efforts will focus on enriched representation, with recruitment via email, social media and Fitbit in-app notifications. At the conclusion of Wave 1, the study investigators will assess study participation rates and revise enrollment targets for Wave 2.

Wave 2 (Augmented Enrollment): Upon the completion of Wave 1, Wave 2 efforts will discontinue enrollment of sufficiently represented groups and focus entirely on increasing participation of under-represented groups, as defined by race and ethnicity (e.g., Caucasian, African American, Asian, Latina/Latino, Native Americans/ Indigenous Populations), biological sex at birth (Female; Male), age (18 - 40; over 40), sexual orientation/gender identity (Heterosexual; LGBTQIA+) and whether participants are Fitbit users. The multi-stakeholder engagement refers to the study's collaboration with major stakeholders such as active groups or KOLs of underrepresented groups of interest to co-develop adaptive recruitment goals and strategies.

The study will aim to recruit a study sample that reflects enriched participation of traditionally underrepresented groups in the US. The Wave 2 targeted recruitment strategy will include five demographic factors, plus ownership of Fitbit devices. If Wave 1 enrollment is not sufficiently at parity with population representation, Wave 2 efforts will target up to an additional 4,000 participants to demographically balance the aggregate sample (potential total n <= 14,000).  Table 1 below outlines the targeted recruitment factors.

Informed Consent

If they choose to enroll, participants will be directed to a consent form administered via an in-app onboarding flow that describes the data that is collected and how it is used to advance the goals of the study. Each prospective study participant must pass the eligibility survey before proceeding to an assurance of understanding flow, where they are given a summary of the summary and key points of the consent. They are then quizzed on this content before they can proceed. There are multiple points to ensure comprehension of the study and consent. Prospective study participants agree to an E-consent, which provides short explanations for the data they are sharing and how they are used and a long form consent which is a full IRB approved consent form detailing the data use and rights. Also, at the top of the form is a text box including basic details of the study, designed to make sure participants receive the most critical details, regardless of how.

The consent form will be associated with a unique participant ID that is stored securely. The purpose of this ID is to be able to link the consent form to the data collected, in order to confirm that data was collected with consent if needed. As mobile sensing is nonintrusive participants may easily forget they are participating in the study, the participants will receive daily notifications reminding them that they are enrolled in the study.

Participants will be informed that if they choose to withdraw, they will have the option to remove or retain all or part of their data. In addition, even if they do not withdraw from the study, they may choose at any time to have any portion of their data omitted, giving the dates and times between which they want their data to be removed.

Each consent form will be read and signed electronically, since this study is entirely online. Participants will indicate their consent to participate in this study by typing their name and email to indicate that they have read and understood the informed consent document. Participants will be given the option to contact research staff if they have any concerns regarding the consent. Participants will be emailed a copy of the consent form to their email address.

Data Collection

*Digital Measures*

The study includes digital measurement from both smartphones and wearable physiological devices. After enrolling in the study, participants will be asked to grant permissions to their devices.

Mobile Data
Patterns of device use will be measured objectively via the study mobile app. Measurements will include:

*Screen on Time and Application Usage.* The amount of screen time derived from smartphone app usage data such as time of day and day of week computed per application type - application usage will be quantified for categories of apps rather than individual apps, and application categories will be determined by those listed on the Google Play Store.

*Mobility & Semantic Location Information.* Various measures of geographic mobility will be discerned from GPS data such as daily count of travel/stop events per person per day, duration of each travel/stop event (min), and proportion of day spent at home, work or some other location (labeled "other").

*Human Activity Recognition from smartphone accelerometer.* Various movement patterns of human activities of interest such as stationary, walking, running, biking, driving will be derived from data generated by smartphone accelerometers without being limited by phone location or orientation.

*Battery & Charging Status*. Smartphone battery status and plug-in/un-plug events.

Wearable Data
We will collect the following data from participants who sign up with Fitbit devices. As access to devices can vary, participation is allowed with any generation of Fitbit devices. Different Fitbit devices have different sensors and algorithmic capabilities and therefore, not all the measures described below will be available for all participants. Minute-level features include accelerometer-based measures of steps and photoplethysmography-based heart rate measurements. Day-level features include accelerometer-based measures of activity (e.g., total steps, number of floors climbed, and number of active zone minutes), sleep stages, duration and quality, minutely heart rate, daily resting heart rate, respiration rate during sleep, heart rate variability (HRV) during sleep, skin temperature during sleep, blood oxygen saturation (SpO2) during sleep, tonic electrodermal activity (EDA).

## Survey Instruments

Participants will complete a battery of self-report questionnaires at the beginning (baseline) and end of the study (post study) as follows: (Baseline only) Demographics questionnaire and Big Five Inventory (BFI-10) (Caprara et al. 1993)(Caprara et al. 1993; Rammstedt and John 2007), Patient Health Questionnaire (PHQ-8) (Caprara et al. 1993; Kroenke et al. 2009), Generalized Anxiety Disorder Scale (GAD-7) (Löwe et al. 2008), Patient Reported Outcomes Measurement Information System (PROMIS) Sleep Disturbance & Sleep Related Impairment short form (Cella et al. 2010; Yu et al. 2011), PROMIS Emotional Support Short Form 4a. (Cohen, Kamarck, and Mermelstein 1983; Bode et al. 2010), Shortened Smartphone Addiction Scale (SAS) (Kwon et al. 2013), Perceived Stress Scale (PSS) (Cohen, Kamarck, and Mermelstein 1983).

## Ecological Momentary Assessments

The first and last week of the study will include intensive ecological momentary assessments (EMAs). During these weeks, 3x per day, participants will be sent 6 identical survey questions per day for 7 days, on a quasi-randomized schedule. The first 5 questions will be "How X do you feel right now?" where X will be replaced with a specific affect descriptor from {happy, calm, anxious, sad, stressed}, and participants will be asked to rate each item. The final question is "Over the last few hours, who have you spent the most time with?" and participants will choose from {alone, friends, family, spouse/partner, co-workers, co-students}. Finally, every day for the entire duration of the study, participants will be asked "In general, how have you been feeling over the last day?" once each morning.

## Compensation

We leverage a raffle design for incentives and compensation. Participants who complete the required elements of the study will be eligible to enter the raffle to receive a $50 gift card. Specifically, the conditions for eligibility were to: 1) consent and enable sensor collection at study start, 2) complete the pre-study assessments, 3) complete a minimum cumulative of seven days (one week) of daily status assessments and to 4) complete the post-study assessments. We will segment the raffles in batches of 1,000 participants, to avoid delaying delivery of the incentive until the completion of the study (which could be quite some time after the first several thousand participants complete the participant duration of 4 weeks). Up to 1,610 participants may receive this incentive over the course of the study, a 11.5% win ratio.

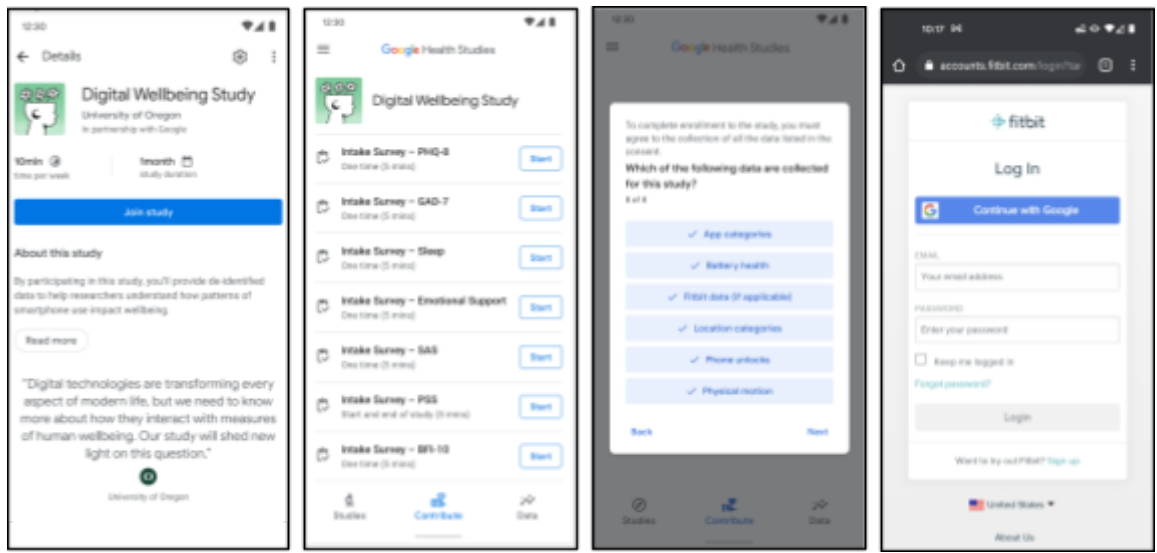

Fig 2. Screenshots of the Android phone application that will be used in the Digital well-being Study. The app facilitates enrollment and consent, baseline and post study surveys, EMAs and logging of the smartphone measures. Participants can optionally link their Fitbit account.

Analysis Plan

The first phase of the hypothesis driven data analysis will be to process the raw data signals to derive interpretable features to test explicit hypotheses. The study will test a range of specific hypotheses and it would be beyond the scope of the current paper to list them all. However, we present an exemplar here to demonstrate how such hypotheses could be tested. For example, one hypothesis that has been proposed in the research literature is that a key mechanism by which device use may affect mental health and well-being is by disrupting healthy sleep behaviors (Alonzo et al. 2021). Specifically, it has been found that device use during the pre-bedtime period may delay sleep onset either by displacing sleep by device use, and/or by increasing psychological arousal prior to lights out, thereby increasing sleep onset latency. In the current study estimates of device use (in the pre-bedtime period will be combined with sleep features, including bedtime and rise time, time in bed, total sleep time, and sleep variability will be derived from phone actigraphy and (in a subset of cases) Fitbit data, to to test the hypothesis that sleep disturbance mediates the relationship between device use and mental health and well-being outcomes. Specifically, we will test whether the hypothesized mechanisms of change (i.e., sleep) mediates the relationship between device use patterns and measures of the change (slope) in measures of mental health and well-being across the study period. Mediating effects will be tested using the strategies outlined by Hayes and Preacher (2010), which are standard in behavioral research trials. Utilizing linear structural equation models, support for mediation will be determined if measures of changes in measures of mental health and well-being become non-significant or are significantly reduced after controlling for the hypothesized mediating variable. We will assess the joint significance of the indirect pathways (i.e., the joint significance of the pathways from the predictor to the mediating variable and from the mediating variable to psychiatric outcomes) using the bias-corrected bootstrap test, and will also employ the bootstrapping technique to test multiple simultaneous mediators (Hayes & Preacher, 2020).

The intensive longitudinal data on these features and self-reports of well-being will be subjected to data reduction techniques such as dynamic factor analysis, which are a set of techniques for estimating common trends in multivariate time series (Zuur et al. 2003). Dynamic factor analysis is a dimension reduction technique that aims to model a multivariate observed time series in terms of a finite number of common trends, with the aim of detecting the smallest number of trends that can summarize the data without losing information. The principle is the same as in other dimension reduction techniques, such as principal component analysis and factor analysis, except that the axes are restricted to be latent smoothing functions over time. The dynamic factor analysis will initially utilize data from the entire 4 week data collection period to examine the patterns of association between daily patterns of device use and daily variations in well-being, and then will go on to examine the data from the two 1-week burst of intensive EMA to examine within-day covariation between patterns of device use and well-being. Variables that do not load on to one of the identified factors, but that are of strong interest as putative risk factors based on theory or previous empirical research, will be investigated in the statistical analyses as univariate predictors.

The second phase of the statistical analyses will consist of a number of analyses to be conducted to address the aims outlined above. A range of analytic techniques will be used including analysis of variance, and specialized multilevel regression techniques that have been used in previous studies of intensive longitudinal data (e.g., [Kuppens, Allen, and Sheeber 2010]), and mixed Markov models (de Haan-Rietdijk et al. 2017). These models are appropriate for testing the hypotheses with our aims, whereby changes in well-being across time can be considered an observed state switch and analyzed using observed Markov models (also referred to as manifest or simple Markov models or Markov chains; (de Haan-Rietdijk et al. 2017; Kapland 2008; Langeheine and van de Pol 2002). These methods can also be used to investigate the temporal relationship between the intensive longitudinal data streams and the optimal time lag between these measurements and state switches (i.e., the time lag between a predictor and the onset of a change in well-being status), which can be investigated by using higher order models, where the current state is modeled as depending on multiple preceding states (Zucchini, MacDonald, and Langrock 2016).

We also plan to investigate the use of neural models to create machine learned representations in our analysis. The intensive longitudinal nature of the data collection will enable large-parameter models to be trained using unsupervised and supervised machine learning techniques. Learning with self-supervision has recently attracted increasing interest as simple pretext tasks can be used to create versatile compressed representations of data. The main advantage of these embeddings is that they provide an initialization for downstream tasks. This is a more label efficient mechanism to train supervised models. The benefits of un- or self-supervised learning techniques studying human behavior is still relatively underexplored. However, the high dimensionality of the and scale of the data described in this protocol lend themselves to this type of analysis. We plan to empirically test whether leveraging these methods has a positive impact on downstream predictive performance.

## Results and Discussion

There is a pressing need for more research into the role that digital device use plays in mental health and well-being, and how these devices may be used to better monitor symptoms and complement existing clinical measures. The Digital Well-being Study is an intensive longitudinal study designed to collect representative US data across a large and representative population to help answer questions about the interaction between digital device use and mental health.

At the time of submission the study infrastructure and app have been designed and built, the institutional review board of the University of Oregon has approved the study protocol and data collection is underway. Data from 4,182 enrolled participants has been collected as of March 27th 2023.

Table 1 shows the distribution of participants enrolled in the digital well-being study at the time of writing. We have been able to achieve a demographic split relatively similar to the 2021 US Census data in terms of race, biological sex, age and sexual orientation.

Figure 3 shows the geographic distribution of the participants based on home state. This map illustrated the geographic representation enabled by performing a study with our mobile platform and ubiquitous wearable and smartphone sensing.

Table 1. Our goals are to recruit a study sample that reflects demographic breakdowns of the US adult population. Statistics as of March 27th 2023.

| FACTOR (Level) | US Census | 2021 US Census | Digital well-being Study |
|---|---|---|---|
| Ethnicity | Hispanic or Latino | 20% | 8% |
|  | Not hispanic or Latino | 80% | 92% |
| Race | Caucasian | 76% | 82% |
|  | Black | 14% | 6% |
|  | Asian | 7% | 3% |
|  | Native American/Alaskan Native | 1% | 1% |
|  | Some other race | 5% | 8% |
| Biological Sex | Male | 48% | 32% |
|  | Female | 48% | 65% |
|  | I Prefer Something Else |  | 3% |
| Age | 18 - < 40 | 55% | 54% |
|  | 40+ | 45% | 46% |
| Orientation | Heterosexual | 88% | 78% |
|  | LGBTQIA+ | 8% | 20% |
|  | I Prefer Something Else | 2% | 2% |
|  | I Don't Know | 2% | - |

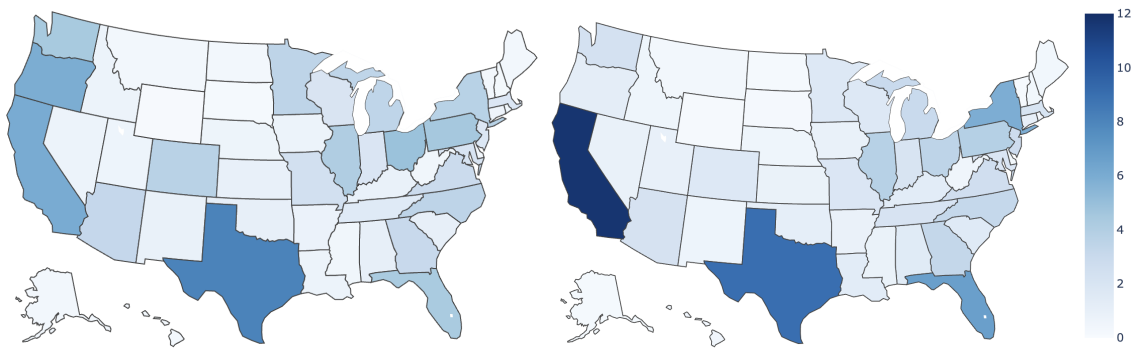

Fig 3. The distribution of participants by home state (in percentage). Left) Digital well-being Study home state density. Right) US 2021 Census population density.

## Limitations

A study of the general population means we cannot draw the same conclusions as a study of a more specific, narrow, group. Unlike studies of medical interns or undergraduate students (Kalmbach et al. 2018; Taylor et al. 2020; Xu et al. 2023) where there may be some common stressors (e.g., start of an internship or an examination period) that might impact all subjects within a similar time frame, our study represents a much more diverse population.

## Conclusions

A large in-situ study could provide valuable insight into the role of digital device use on mental health and well-being. Through the Digital Well-being Study, we intend to answer critical questions based on a large population analysis. We plan to combine measurements from wearable devices, smartphones, surveys and EMAs. By leveraging ubiquitously available devices we aim to sample a more representative snapshot of the US population than has previously been reported. .

## Acknowledgements

We would like to thank the Google Health Studies Team: Abhyudai Burla, Aliza Hoffman, Ana Krulec, Austin Kauble, Chris Geiershouse, Dhruvika Sahni, Dorsa Vahebi Wiseman, Ed Lee, Genevieve Foti, Jamie Flores, Jon Morgan, Karan Goel, Laxmi Kambli, Michael Rule, Nina Shih, Neil Smith, Nupur Banerjee, Pei Zhong, Shalini Sharma, Shelagh McLellan, Tejas Khorana, Varsha Venkat, Zach Wasson.

## Conflicts of Interest

D. McDuff, A. Barakat, A. Jiang, F. Cordeiro and J. Hernandez are employees of Google LLC

## Abbreviations

BFI: Big Five Inventory
EDA: Electrodermal activity
EMA: Ecological momentary assessments
GAD: Generalized Anxiety Disorder
HRV: Heart rate variability
JMIR: Journal of Medical Internet Research
PHQ: Patient Health Questionnaire
PROMIS: Patient Reported Outcomes Measurement Information System
PSS: Perceived Stress Scale
RCT: randomized controlled trial
SAS: Smartphone Addiction Scale

## Authorship Contributions

J. Hernandez, A. Barakat, and N. Allen conceived the research question and initiated the project.

N. Allen, J. Hernandez, and A. Barakat designed the research and wrote the protocol.

D. McDuff wrote the first draft of the manuscript.

All authors revised the manuscript.

N.Allen and J. Hernandez provided senior supervision of the project.

Corresponding author: Daniel McDuff